\documentstyle[12pt]{article}

\newcommand{\be}{\small\begin{equation}}
\newcommand{\ee}{\end{equation}\normalsize\vspace*{-0.1ex}}
\newcommand{\bea}{\small\begin{eqnarray}}

\newcommand{\eea}{\end{eqnarray}\normalsize\vspace*{-0.1ex}}
\newcommand{\bdm}{\small\begin{displaymath}}
\newcommand{\edm}{\end{displaymath}\normalsize\vspace*{-0.1ex}}
\newcommand{\beas}{\small\begin{eqnarray*}}

\newcommand{\eeas}{\end{eqnarray*}\normalsize\vspace*{-0.1ex}}

\newcommand{\n}{\noindent}

\begin{document}


\thispagestyle{empty}
\renewcommand{\thefootnote}{\fnsymbol{footnote}}

\setcounter{page}{0}
\begin{flushright} DESY 94--200\\
UM-TH-94-37\\
hep-ph/9411229\\
November 1994  \end{flushright}

\begin{center}
\vspace*{0.5cm}
{\Large\bf
Naive nonabelianization and resummation of fermion bubble
chains
}\\
\vspace{1.4cm}
{\sc M.~Beneke} \\
\vspace*{0.3cm} {\it Randall Laboratory of Physics\\
University of Michigan\\ Ann Arbor, Michigan 48109, U.S.A.}\\[0.6cm]
and\\[0.6cm]
{\sc V.~M.~Braun\footnote{On leave of absence from
St.Petersburg Nuclear Physics Institute, 188350 Gatchina,
Russia}} \\ \vspace*{0.3cm} {\it DESY\\ Notkestr. 85\\ D--22603
Hamburg, Germany}\\[1.4cm]
{\bf Abstract}\\[0.3cm]
\end{center}

\parbox[t]{\textwidth}{
We propose to extend the Brodsky-Lepage-Mackenzie scale-fixing
prescription by resumming exactly any number of one-loop vacuum polarization
insertions into one-loop diagrams. In this way,  one makes maximal use
of the information contained in one-loop perturbative corrections
combined with the one-loop running of the effective coupling. The
scale ambiguity at leading order is converted into an intrinsic uncertainty
of perturbative approximations induced by IR renormalons.
Practical implementation of this resummation requires only
knowledge of one-loop radiative corrections with non-vanishing gluon mass.
We find that higher order corrections to the pole
mass and the top quark decay width are dominated by renormalons
already in low orders
and demonstrate the impact of eliminating the pole
mass on the convergence of the perturbative series.
}
\vspace*{0.3cm}
\begin{center}
{\it submitted to Phys. Lett. B}
\end{center}

\newpage
\renewcommand{\thefootnote}{\arabic{footnote}}
\setcounter{footnote}{0}


{\bf 1.} With experiments growing ever more accurate, the problem of
scheme- and scale-fixing in truncated perturbative expansions for
observables

\be\label{quantity}
R -R_{\rm tree}= \alpha(Q) \sum_{n=0}^N r_n \alpha(Q)^n
\ee

\n in QCD continues to be of acute practical interest. Inherent to any
scale-fixing
procedure is a guess of uncalculated higher order corrections
and the impossibility of assessing with rigour the quality of such a
guess for any particular observable. Yet some prescriptions may
be closer to general expectations. Such a prescription has been
formulated by Brodsky, Lepage and Mackenzie (BLM) \cite{BRO83}. Motivated
by the observation that in QED all effects of the running coupling are
associated with the photon vacuum polarization, they suggested to
absorb the contribution of one fermion loop insertion into lowest
order diagrams in the scale of the lowest-order correction. Thus,

\bea
&& r_0\alpha(Q)+[r_{10}+r_{11} N_f]\alpha(Q)^2+\ldots \equiv
r_0\Big\{\alpha(Q)+[\delta_1+(-\beta_0)\,d_1]\alpha(Q)^2+\ldots\Big\}
\nonumber\\
&& \hspace*{1cm}
\equiv\,r_0\Big\{\alpha(Q^*_1)+\delta_1\alpha(Q)^2+\ldots\Big\}
\eea

\n where $\beta_0=
(-1/(4\pi)) (11-(2 N_f)/3)$ is the first coefficient of the QCD
beta-function, and the new scale is given by $Q^*_1 =Q\,e^{-d_1/2}$.
In QCD the main effect of this scale redefinition is not to absorb
a relatively small contribution proprotional to $N_f$ from the fermion
loop itself, but to absorb contributions of other diagrams as well,
via the substituion $N_f\rightarrow N_f-33/2$.

In many cases, such as observables derived
from the hadronic vacuum polarization \cite{MAX} and the pole mass,
it turns out
that $\delta_1$ is small and $r_0\alpha(Q^*_1)$ is in fact a very good
approximation to the exact two-loop result.
This suggests that often the bulk
contribution to radiative corrections in QCD can be obtained from diagrams
with fermion loop insertions, after the full $\beta_0$ of QCD is restored,
a procedure which we shall refer to as naive nonabelianization\footnote{
This denomination has been introduced by D.~Broadhurst
(private communication).}.
Beyond this empirical fact,
additional justification for this extrapolation
derives from the observation that in large orders, the series of
radiative corrections is expected to grow as $n!\,(-\beta_0)^n\alpha(Q)^n$
from insertion of $n$ fermion loops and dominance
of these graphs (plus naive nonabelianization) can persist to relatively low
orders. At the same time the fast growth, $\propto n!$, indicates that
the effect of including more than one fermion loop insertion into the
definition of $Q^*$ can be numerically significant.

In this note, we investigate the effect of multiple one fermion loop
insertions and provide a generalization of the leading order BLM scale
$Q^*_1$ by summing exactly to all orders all fermion loop insertions in
lowest order gluon exchange diagrams.
 This corresponds to defining $\alpha(Q^*)$ as the
average of the one-loop running coupling over the lowest order radiative
corrections of a particular process. This proposal can hardly be considered
new \cite{LEP93},
though it has never been pursued in the literature, apparently because
of lack of its
computational implementation. We shall show that the resummation of this class
of diagrams requires as input only lowest order radiative corrections
computed with a finite gluon mass. Resumming fermion loop insertions has at
least three merits:
\begin{itemize}
\item It makes maximal use of the information contained in the one-loop
radiative corrections combined with one-loop running of the coupling and
is thus the logical realization of the BLM prescription at one loop.
\item Using naive nonabelianization it absorbs a class of supposedly
large corrections into the normalization of the lowest order coupling.
Numerically, this turns out to be most relevant in heavy quark physics.
\item It affords a semiquantitative estimate of the ultimate accuracy
of perturbative approximations due to renormalon divergence.
\end{itemize}
We would like to note that this resummation of a subclass of diagrams which is
not based on any systematic parameter is delicate in the sense that it can
only be justified a posteriori, akin to, say, tadpole improvement in
lattice perturbation theory \cite{LEP93} or
resummation of ``Sudakov-$\pi^2$'s''.
It is for this reason that we prefer the
discussion in the context of scale-fixing methods, though this is irrelevant
for practical purposes.

In the following section, we introduce our notation and formulate the
resummation prescription in precise terms. In section 3, we collect the
techniques necessary to calculate any number of fermion loop insertions and
their sum. The final section presents an application to the pole mass
and the top quark width. This example demonstrates the numerical importance
of higher order loop insertions and the impact of eliminating the pole
mass from heavy quark decay widths on the convergence of perturbation
series even in low orders. A more detailed account of techniques as well
as their application to the phenomenologically most relevant case of
beauty and charm decays will be presented in a separate
publication \cite{BBB94}.\\


{\bf 2.} Our focus is on renormalization group invariant quantities,
which are dominated by a large scale Q, such as in Eq.~(\ref{quantity}).
We shall also assume that lowest order radiative corrections $r_0$ are
given by one gluon exchange and do not involve the gluon self-coupling.
We fix a renormalization scheme that does not introduce artificial
$N_f$-dependence, for instance $\overline{\rm MS}$. Then $(n+1)$-th order
corrections can be written as

\be r_n=r_{n0}+r_{n1} N_f+\ldots+r_{nn} N_f^n,
\ee

\n where $r_{nn}$ originates from $n$ fermion-loop insertions into the
lowest order radiative corrections. We rewrite $r_n$ as

\be r_0\left[\delta_n+(-\beta_0)^n d_n\right],
\ee

\n where $d_n$ is  determined by the requirement that it absorbs the largest
power of $N_f$. In particular,
$d_n=(-6\pi)^n r_{nn}$.
Again we note that through $\beta_0$,
the second term in brackets absorbs the contributions from
diagrams with less than $n$
fermion loops associated with the gluonic part of the one-loop running of
the coupling. It will be absorbed into the scale of the lowest order
correction. To this end, for the perturbative expansion, truncated at
order $N+1$, we introduce

\be\label{M}
M_N(-\beta_0\alpha(Q)) = 1+\sum_{n=1}^N d_n (-\beta_0\alpha(Q))^n
\ee

\n as a measure of how much the lowest order correction is modified by
summing $N$ one-loop vacuum polarization insertions and define

\be \label{scale} \alpha(Q^*_N) = \alpha(Q)\,M_N(a)\, \qquad
a\equiv-\beta_0\alpha(Q).
\ee

\n With these definitions

\be R-R_{\rm tree} = r_0\left(\alpha(Q^*_N)+\sum_{n=1}^N\delta_n\alpha(Q)^
{n+1}\right),
\ee

\n where the sum contains the ``genuine'' higher order corrections, not
related to the scale dependence of the coupling\footnote{In the present
form, the $\delta_n$ still contain two and higher loop vacuum polarization
insertions, which upon incorporation of higher loop running of the coupling
would also be absorbed into $Q^*_N$, see ref.~\cite{BRO94} for
the case $N=2$.}.

We remark that for $N>1$, the solution $Q^*_N$ to
Eq.~(\ref{scale}) depends on the value of $\alpha(Q)$. The coupling
dependence of any generalization of the BLM prescription beyond leading
order has previously been noted in \cite{GRU92,BRO94}.  For $N=\infty$,
$\alpha(Q^*)$ ($Q^*\equiv Q^*_\infty$) is transparently interpreted as the
running coupling averaged over the lowest order corrections. Isolating the
integration over gluon virtuality in the lowest order correction (ignoring
renormalization for now), we may write

\be r_0\alpha(Q)=\alpha(Q)\int\mbox{d}^4 k\,F(k,Q)\,\frac{1}{k^2} .
\ee

\n Then

\be \label{average}
r_0\alpha(Q^*) = \frac{r_0\alpha(Q)}{1-\beta_0\alpha(Q)\ln({Q^*}^2/Q^2)}
= \int\mbox{d}^4 k\,F(k,Q)\,\frac{\alpha\left(
k\exp[C/2]\right)}{k^2},
\ee

\n where the scheme-dependent constant $C$ is the constant part of
the fermion loop, see below.
The integral in Eq.~(\ref{average})
is not well defined  due to the Landau pole of the one-loop
running coupling. Equivalently $\lim_{N\rightarrow\infty}M_N$ does
not exist, because the $d_n$ exhibit the familiar renormalon divergence,
$\propto n!$. However, assuming perturbative series are asymptotic,
$Q^*$ can still be defined -- see below -- up to terms suppressed by
a power of the large scale $Q$. After resummation $\alpha(Q^*)$
is formally independent of the finite renormalization $C$ for the
fermion loop. All scheme-dependence introduced by one-loop counterterms
is eliminated and the accuracy of perturbative predictions is
limited only by the divergence of the series and scheme-dependence
in unknown genuine higher order corrections.

It is worth noting that the resummation of vacuum polarization insertions
could be extended to two (and consecutively higher) loop
insertions by including
the effect of two-loop evolution of the coupling into $Q^*$ and
introducing a second scale $Q^{**}$ for the genuine two-loop correction
plus an infinite number of one-fermion-loop insertions into these,
see ref.~\cite{BRO94} for one insertion. The introduction of new
BLM scales at each order closely parallels the $1/N_f$-expansion,
employed in the analysis of renormalon singularities. Since it is known
that the nature as well as normalization of renormalon singularities
is not correctly provided by the leading term in the $1/N_f$-expansion,
one might therefore question the usefulness of the restriction to $Q^*$
alone. However, contributions of higher order in $1/N_f$ to the
normalization of renormalons and therefore the estimate of ambiguity
of perturbative approximations, start at increasingly larger number
of loops. Thus, in all phenomenologically interesting cases, where the
divergence of the series starts at comparatively low orders in
perturbation theory, the incalculability of the normalization is
practically irrelevant and the ambiguity from resumming multiple
one-fermion loop insertions should provide a good guide to the limits
of perturbation theory.\\


{\bf 3.} To compute the coefficient $r_n$ with $n$ fermion loop
insertions (``$n$ bubbles'') into lowest order radiative corrections,
it is often useful to calculate the Borel transform

\be\label{borel}
B[R](u) = \sum_n \frac{r_n}{n!}(-\beta_0)^{-n} u^n
\ee

\n directly and use it as a generating function for the coefficients
$r_n$ \cite{BEN93}:

\be\label{genfunction}
 r_n = (-\beta_0)^n \frac{d^n}{du^n} B[R](u)_{|_{u=0}}.
\ee

\n In many cases, in particular observables
involving more than one mass scale, the exact Borel transform is
difficult to obtain or even if it is obtainable, taking derivatives
is not always a simple task. We can avoid this complication by
exploiting that the information about $n$ bubble coefficients is
contained in the lowest order radiative corrections, calculated
with non-vanishing gluon mass $\lambda$ (in Landau gauge):

\be\label{finitemass}
r_0(\lambda^2)= \int d^4k\, F(k,Q)\frac{1}{k^2-\lambda^2}\,.
\ee

\n Indeed it has been
emphasized \cite{BBZ94} that the residues of IR renormalon poles are
given by nonanalytic terms in $\lambda^2$ in a small-$\lambda^2$
expansion, and in ref.~\cite{SV94} the insertion of one bubble has been
expressed as an integral over $\lambda^2$. We fill the gap between one
bubble and the asymptotic behaviour and provide a representation for
$r_n$ and the sum of contributions with any number of one-fermion loop
insertions (``bubble sum'') in terms of the lowest order radiative
correction with finite gluon mass as only input. We restrict attention
to observables in Euclidean space or such that can be obtained
upon analytic continuation from Euclidean space. We shall also assume
that no explicit renormalization is needed except for the fermion loop
insertions. This assumption can easily be relaxed \cite{BBB94}. It is
convenient to use the Landau gauge, the final result being of
course gauge-independent.

We start with the ``bubble sum''. The gluon propagator with summation of
an arbitrary number of bubbles can be written as (in Landau gauge)

\be\label{propagator}
D^{AB}(k) = i\delta^{AB}\frac{k_\mu k_\nu-k^2 g_{\mu\nu}}{k^4}
\frac{1}{1+\Pi(k^2)}
\ee

\n where

\be
\Pi(k^2) = a \ln\left(\frac{-k^2}{Q^2}e^C\right)
\ee

\n
For a generic contribution to a certain physical amplidude with
Euclidian external momenta, one can separate the integration over the
gluon momenta to write the contribution of the bubble sum as
(cf. Eq.~(\ref{average}))

\be\label{bs}
r_0 \alpha(Q) M(a)
=\int d^4k\, F(k,Q)\frac{1}{k^2}\frac{\alpha(Q)}{1+\Pi(k^2)} .
\ee

\n where the transverse projector that appears in the gluon propagator
in Eq.~(\ref{propagator}) is assumed to be included in the function $F(k,Q)$.
The next step is to substitute $(1+\Pi(k^2))^{-1}$ by the dispersion
relation

\be
\frac{1}{1+\Pi(k^2)} = \frac{1}{\pi}\int_0^\infty d\lambda^2\,
\frac{1}{k^2-\lambda^2} \frac{{\rm Im}\Pi(\lambda^2)}{|1+\Pi(\lambda^2)|^2}
+\int_{-\infty}^\infty d\lambda^2\,\frac{1}{k^2-\lambda^2}
\frac{\lambda_L^2}{a} \delta(\lambda^2-\lambda_L^2)
\ee

\n where

\be
\lambda_L^2 =- Q^2\exp[-1/a-C]
\ee

\n is the position of the Landau pole. Further writing

\be
\frac{1}{k^2-\lambda^2} =\frac{1}{\lambda^2}\left(
\frac{k^2}{k^2-\lambda^2} -1\right),
\ee

\n interchanging the order of integrations in $k$ and $\lambda^2$,
we arrive at

\be
r_0 M(a) = -\int_{-\infty}^\infty\frac{d\lambda^2}{\lambda^2}
\,\left\{\frac{a \,\theta(\lambda^2)}{|1+\Pi(\lambda^2)|^2}-
\frac{\lambda^2_L}{a}\delta(\lambda^2-\lambda_L^2)\right\}
\Big[r_0(\lambda^2)-r_0(0)\Big]\,.
\ee

\n Since by assumption the integral over gluon momentum in
Eq.~(\ref{finitemass}) is ultraviolet finite, $r_0(\lambda^2)$
decreases to zero at $\lambda\to\infty$. Thus, integrating
by parts, we finally get

\be\label{rBS}
r_0 M(a) = \int_0^\infty d\lambda^2\, \Phi(\lambda)\,r'_0(\lambda^2)
+\frac{1}{a}[r_0(\lambda_L^2)-r_0(0)]
\ee

\n where $r'_0(\lambda^2) \equiv (d/d\lambda^2) r_0(\lambda^2)$ and

\be\label{Phi}
\Phi(\lambda)= -\frac{1}{a\pi}\arctan\left[\frac{a\pi}
{1+a\ln(\lambda^2/Q^2 e^C)}\right] -\frac{1}{a}
\,\theta(-\lambda_L^2-\lambda^2)\,.
\ee

\n Note that the term with the $\theta$-function exactly cancels the
jump of the $\arctan$ at $\lambda^2=-\lambda_L^2$.

Eq.~(\ref{rBS}) presents the desired answer for the sum of
diagrams with any number of fermion bubbles in terms of an integral
over gluon mass. This result has a very transparent structure: the
quantity $r'_0(\lambda^2)$ (with certain reservations) can be considered
as the contribution
to the integral from gluons of virtuality of order $\lambda^2$, and
the function $\Phi(\lambda)$ can be understood as an effective charge.
At large scales  $\Phi(\lambda)$ essentially coincides with
 $\alpha_v$, the QCD coupling in so-called $V$-scheme
\cite{BRO83}, but in difference to
it remains finite at small $\lambda$. The absence of a Landau pole
in this effective coupling exhibits another welcome feature of
Eq.~(\ref{rBS}). The integral is a well-defined number and the fact
that we have started with an ill-defined expression in Eq.~(\ref{bs})
due to the Landau pole (equivalently, attempted to sum a non-Borel
summable series) is isolated in the Landau pole contribution
$r_0(\lambda_L^2)$ . Whenever IR renormalons are present, $r_0(\lambda^2)$
develops a cut at negative $\lambda^2$. One can show that the
real part of the above prescription for the bubble sum coincides
with the principal value of the Borel integral \cite{BBB94} and,
in particular, coincides with the Borel sum, when it exists. The
imaginary part provided by $r_0(\lambda_L^2)$ coincides with the
the imaginary part of the Borel integral, when the contour is deformed
above (or below) the positive real axis. We therefore adopt the
real part of Eq.~(\ref{rBS}) as a defnition of the bubble sum
contribution and include the imaginary part (divided by $\pi$)
as an estimate of intrinsic
uncertainty from resumming a (non-Borel summable) divergent series
without any additional nonperturbative input. Note that this
imaginary part is proportional to a power of $\lambda^2_L/Q^2\sim
\exp(-1/a(Q))$ and therefore suppressed by powers of $\Lambda_{\rm QCD}/
Q$.

We remark that Eq.~(\ref{rBS}) applies without
modification to quantities like inclusive decay rates, which can be
obtained starting from a suitable amplitude in Euclidian space and
taking the total imaginary part upon analytic continuation
to Minkowski space. The structure of the $\lambda^2$-integral remains
unaffected, and it is only the quantity $r'_0(\lambda^2)$ which should be
substituted by the corresponding decay rate calculated with finite
gluon mass (in addition, no explicit renormalization is needed, when
the decay rates are expressed in terms of pole masses).

To obtain the coefficient $r_n$ with $n$ bubble insertions it suffices
to find a representation for the Borel transform, see
Eq.~(\ref{genfunction}). We relax the requirement of no explicit
renormalization, but assume at most a logarithmic ultraviolet
divergence. Regularizing in $d=4-2\epsilon$ dimensions,
the {\em bare} lowest order correction calculated with the finite
gluon mass has
the following asymptotic form:

\be
r_0^{bare}(\lambda^2,\epsilon) \stackrel{\lambda\to\infty}{=}
-\frac{r_\infty(\epsilon)}{\epsilon}
\left(\frac{Q^2}{\lambda^2}\right)^\epsilon.
\ee

\n Inverting

\be
r_0(\lambda^2) =\frac{1}{2\pi i}\int_C du\, \Gamma(-u)\Gamma(1+u)
\left(\frac{\lambda^2}{Q^2}e^C\right)^u B[R](u)
\ee

\n from ref.~\cite{BBZ94} and using the expressions for
renormalization of the Borel transform in App.~A of ref.~\cite{BB94},
we find \cite{BBB94}

\be\label{final}
B[R](u) = -\frac{\sin(\pi u)}{\pi u }\int_0^\infty d\lambda^2\,
\left(\frac{\lambda^2}{Q^2}e^C\right)^{-u}
\left[ r'_0(\lambda^2)-\frac{r_\infty}{\lambda^2}
\theta(\lambda^2-Q^2)\right] + R(u)
\ee

\n where

\be
R(u) = \frac{1}{u}\left(\tilde G_0(u)-\frac{\sin(\pi u)}
{\pi u} r_\infty e^{-uC}\right)\,,
\ee
\be
G_0(u) \equiv \sum_{n=0}^\infty g_n u^n
= \frac{1}{(4\pi)^{-u}}\frac{\sin(\pi u)}{\pi u}
\frac{\Gamma(4+2u)}{6\Gamma(1-u)\Gamma(2+u)^2}\,r_\infty(u)\,,
\ee
\be
\tilde G_0(u) = \sum_{n=0}\frac{g_n}{n!}u^n\,,
\ee

\n and $r_\infty\equiv r_\infty(0)$. Thus $r_n$ can basically be
expressed in terms of the integrals

\be
\int_0^\infty d\lambda^2\,\ln^k(\lambda^2/Q^2)\,r'_0(\lambda^2)
\qquad k\le n\,.
\ee
\bigskip


{\bf 4.} As a first application, we consider the relation between
the pole and the $\overline{\rm MS}$ mass of a heavy quark,
defined as

\be\label{pole}
m_{\rm pole} =m_{\overline{\rm MS}}(m)\left[1+\frac{C_F\alpha}{4\pi}
\sum_{n=0}^\infty r_n
(-\beta_0\alpha(m))^n\right]
\ee

\n The one-loop mass shift with finite gluon mass, $r_0(\lambda^2)$,
equals ($x\equiv \lambda^2/m^2$):

\be
r_0(\lambda^2) = 4+x-\frac{x^2}{2}\ln x-
\frac{\sqrt{x}(8+2x-x^2)}{\sqrt{4-x}}
\Bigg\{\arctan\left[\frac{2-x}{\sqrt{x(4-x)}}\right]+
\arctan\left[\frac{\sqrt{x}}{\sqrt{4-x}}\right]\Bigg\}
\ee

\n Using the formulae collected in sect.~3, we easily
compute the contributions of finite number of bubbles,
collected in the second column\footnote{In this case the
exact Borel transform is simple \cite{BB94}. We have
checked that the coefficients obtained from
differentiating Eq.~(\ref{final}) agree with the
derivatives of the exact expression of ref.~\cite{BB94}.}
of Table~1.
The third column contains coefficients for $n$ fermion loop
insertions
into one-loop radiative corrections to the top decay
width (in the limit $m_t\gg m_W$),

\be\label{topwidth}
\Gamma = \frac{G_F}{\sqrt{2}}\frac{m_t^3}{8\pi}
\left[1+2\frac{C_F\alpha}{4\pi} \sum_{n=0}^\infty g_n
(-\beta_0\alpha(m_t))^n\right]
\ee

\n The one-loop correction with finite gluon mass
$g_0(\lambda^2)$ has been taken from ref.~\cite{SV94} and $m_t$
is the pole mass of the top. The coupling $\alpha$ is always
taken in the $\overline{\rm MS}$ scheme ($C_{\overline{\rm MS}}=
-5/3$).
For both, the pole mass and the top decay width, the coefficients
grow very rapidly, and roughly in the same proportion:

\be
r_n/r_{n-1} \sim g_n/g_{n-1} \sim 2n
\ee

\n This result can be expected because the asymptotic behaviour
of the perturbation series in high orders in both cases is governed
by an infrared renormalon $r_n\sim n!(1/u)^n \alpha(m)^n$ with $u=1/2$
\cite{BB94,BIG94}. It is remarkable that the numerical values are
close to their asymptotic ones already in low orders.

In particular, the growth of coefficients for the top decay
width is completely due to the parametriziation in terms of the
pole mass. It has already been conjectured on the evidence of cancellation
of the leading IR renormalon singularities \cite{BBZ94,BIG94} that
radiative corrections to heavy particle decays are strongly reduced
in high orders if the pole mass is eliminated in favour of a mass
parameter
defined at short distances, e.g. the $\overline{\rm MS}$ mass.
That this phenomenon is already relevant to low orders is
clearly displayed by the fourth column of Table~1, where
the coefficients are given, when Eq.~(\ref{topwidth}) is
expressed in terms of the $\overline{\rm MS}$ mass: the size
of coefficients $g_n^{\overline{\rm MS}} =g_n+(3/2)r_n$ is drastically
reduced. In very large orders, the coefficients are now
sign-alternating and are governed
by an ultraviolet renormalon at $u=-1$. The odd pattern of size of
coefficients comes from interplay of an infrared renormalon
at $u=3/2$ and the ultraviolet renormalon at $u=-1$, which finally
takes over at $n\approx 7$. Note, that
due to the factor $\exp[- u C_{\overline{MS}}]=
\exp(5/3 u)$ in the normalization of renormalons, the
$\overline{\rm MS}$
scheme strongly favours IR renormalon dominance. For this reason,
quantities which have a leading infrared renormalon  (at $u=1/2$ above)
approach the asymptotic regime at comparatively low orders,
wheras onset of the asymptotic regime is delayed, when UV renormalons are
leading \cite{BEN93b}.

In Table~2 we examine the modification of the
lowest order radiative
correction through summation of N fermion loops. We show the
factors $M_N(a)$ defined in Eq. (\ref{M}) and the values
of the BLM scales $Q^*_1$ and $Q^*_\infty $ (see Eq.~(\ref{scale}))

\be\label{compare}
   Q^*_1 = \exp\left[-\frac{1}{2a}(M_1(a)-1)\right] \qquad
   Q^*_\infty = \exp\left[-\frac{1}{2a}
   \left(1-\frac{1}{M(a)}\right)\right]\,,
\ee

\n taking
representative values $\alpha(m_c)=0.35$,
$\alpha(m_b) =0.2$, $\alpha(m_t)=0.1$.
The first two columns display results for the c-quark and b-quark pole
masses, respectively. We conclude that at the charm scale, the relation
between the pole and $\overline{\rm MS}$ cannot be improved
beyond a two-loop calculation.
In the third column we give the results
for the top decay width. In the last two columns we extrapolate
the result for the top width to the scale $m_b\approx 4.8\,$GeV
and show partial sums before and after elimination of the pole
mass. This corresponds to b-quark semileptonic decay width with
zero invariant mass transferred to leptons and gives an anticipation of
the size of radiative corrections to be expected for semileptoinic
decays \cite{BBB94}. In both cases -- with and without pole mass --
the BLM scale turns out to be rather low. The important point is
that a low {\it value} of the BLM scale alone does not indicate a
breakdown of perturbation theory. The failure of perturbation theory
due to its ultimate divergence is indicated by the {\it uncertainty}
in $Q^*$ (or, equivalently, $M(a)$, see Table~2), estimated by the
imaginary part
of the Landau pole
contribution to Eq.~(\ref{rBS}):

\be
        \delta M(a) = \frac{1}{\pi a r_0}{\rm Im}\,r_0(\lambda_L^2)
\ee

\n
Note that in fact there is no convincing argument to include the factor
$1/\pi$ in this estimate, except that upon inspection of Table~2 we find
this estimate closer to the estimate of uncertainty from the minimal
term in the perturbative series. This ambiguity, and also the fact that
the nature of the
renormalon singularity is not determined correctly in the bubble sum
approximation,
indicates that the given error is only a semiquantitative estimate.

Note that $Q^*_\infty$ can be {\em larger} than $Q^*_1$ because
the one-loop correction $M(a)= 1+d_1 a +\ldots$ is always large, and
keeping the first term only in the expansion of $M(a)$ in the denominator
of the second expression in Eq.~(\ref{compare}) --- which is the only
consistent way
to compute $Q^*_1$ --- is in fact a bad approximation. Also note that
the BLM scale $Q_1^*$ after elimination of the pole mass in favour
of the $\overline{\rm MS}$ mass in the decay width is much smaller
than the BLM scale computed with the pole mass, although the
individual coefficients
in the perturbative expansion are reduced, see Table~1.
This highlights once more than the low value of the BLM scale
by itself has no meaning
with respect to failure of the perturbation theory.\\


{\bf 5.}
To conclude, we have proposed to extend the
BLM scale-fixing prescription
by resumming exactly all one-loop vacuum polarization insertions in
lowest order radiative corrections, and have worked out a technical
framework for
the implementation of this program. This generalization is natural,
in the spirit of BLM,
building a bridge between the problem of setting the scale in low
orders and divergences of perturbative expansions in large orders,
but delicate in a sense that resummation of a particular subclass of
diagrams combined with naive nonabelianization is not based
on any systematic parameter.
We find that renormalon ambiguities in summation of the perturbative
series are translated to uncertainties of the BLM scales.
In estimating the overall uncertainty of the calculation, these
must be combined with the uncertainty from unknown genuine corrections in
higher orders. If naive nonabelianization works --- which can
only be justified {\em a posteriori} --- the latter may even be smaller
than the uncertainty due to infrared renormalons.
The renormalon uncertainties in BLM scales, rather than small
values of the scales themselves,
indicate the ultimate accuracy of perturbation theory.

The few examples considered above indicate that this procedure
can be most fruitful for heavy quark physics, where we observe that the
behaviour of perturbative series is consistent
with the asymptotic expansion already in low orders.
Thus, matching explicit calculations of
lower order corrections with asymptotic formulae one can expect
to get a significant benefit.
Immediate phenomenological applications include inclusive B-decays, in
which case the one-loop fermion loop insertion has been calculated in
\cite{LSW94}, and will be considered elsewhere \cite{BBB94}.

\bigskip

{\bf Acknowledgements.} M.~B. would like to thank Chris Maxwell
and Ira Rothstein for discussions.


\newpage


\newpage
\begin{table}[h]
\addtolength{\arraycolsep}{0.2cm}
$$
\begin{array}{|l||c|c|c|}
\hline
n & r_n & g_n & g_n^{\overline{\rm MS}}\\ \hline\hline
0 & 4 & (-2\pi^2)/3+5/2 & (-2\pi^2)/3+17/2 \\
1 & 71/8+\pi^2 & -17.15684 & 10.96007 \\
2 & 70.490601 & -94.02309 & 11.7128 \\
3 & 439.43538 & -626.0250 & 33.128 \\
4 & 3495.6957 & -5208.637 & 34.907 \\
5 & 35358.744 & -52654.86 & 383.3 \\
6 & 423257.13 & -634891.3 & -5 \\
7 & 5939873.7 & -8898590 & 1.05\times 10^4 \\
8 & 94962946 & -1.42473\times 10^7 & -2.83\times 10^4\\ \hline
\end{array}
$$
\caption{Coefficients for $n$ fermion loop insertions into one-loop
radiative corrections for the pole mass and the top decay width normalized
to the pole and $\overline{\rm MS}$ mass.}
\label{tab1}
\end{table}

\begin{table}[h]
$$
\begin{array}{|c||c|c|c|c|c|}
\hline
N & \mbox{c-quark mass} & \mbox{b-quark mass} & \mbox{I} &
\mbox{II} & \mbox{III} \\ \hline\hline
0 & 1 & 1 & 1 & 1 & 1 \\
1 & 2.176 & 1.623 & 1.252 & 1.559 & 1.759 \\
2 & 3.286 & 1.935 & 1.335 & 1.967 & 1.867 \\
3 & 5.024 & 2.193 & 1.368 & 2.328 & 1.908 \\
4 & 8.492 & 2.467 & 1.385 & 2.727 & 1.913 \\
5 & 17.30 & 2.835 & 1.395 & 3.265 & 1.922 \\
6 & 43.76 & 3.420 & 1.402 & 4.126 & 1.922 \\
7 & 137.0 & 4.514 & 1.408 & 5.732 & 1.926 \\
8 & 511.0 & 6.838 & 1.414 & 9.151 & 1.924 \\ \hline
\infty & 1.712\pm 0.608 & 2.041\pm 0.201 & 1.408\pm 0.068
& 2.094\pm 0.295 & 1.928\pm 0.001 \\ \hline
Q^*_1 & 0.096 \,m_c & 0.096 \,m_b & 0.122 \,m_t
& 0.122 \,m_b & 0.058 \,m_b \\
Q^*_\infty & 0.437 \,m_c & 0.147 \,m_b & 0.089 \,m_t & 0.140 \,m_b
& 0.164 \,m_b \\
\hline
\end{array}
$$
\caption{Modification $M_N(a)$ of the lowest order correction through
summation of $N$ fermion loops. For charm, we have taken $a=0.251$,
for beauty $a=0.133$ and for top $a=0.06$. Column I is for top quark
width with pole mass at $a=0.06$, columns II and III extrapolate
to $a=0.133$ before (II) and after elimination of the pole mass (III).}
\label{tab2}
\end{table}

\end{document}